\documentclass[]{spie}  
\usepackage[]{graphicx}

\title{The Keck Array: a pulse tube cooled CMB polarimeter} 

\author{C.~D.~Sheehy,\supit{a,b}
P.~A.~R.~Ade,\supit{c}
R.~W.~Aikin,\supit{d}
M.~Amiri,\supit{e}
S.~Benton,\supit{f}
C.~Bischoff,\supit{g}
J.~J.~Bock,\supit{d,h}
J.~A.~Bonetti,\supit{h}
J.~A.~Brevik,\supit{d}
B.~Burger,\supit{e}
C.~D.~Dowell,\supit{d,h}
L.~Duband,\supit{i}
J.~P.~Filippini,\supit{d}
S.~R.~Golwala,\supit{d}
M.~Halpern,\supit{e}
M.~Hasselfield,\supit{e}
G.~Hilton,\supit{j}
V.~V.~Hristov,\supit{d}
K.~Irwin,\supit{j}
J.~P.~Kaufman,\supit{k}
B.~G.~Keating,\supit{k}
J.~M.~Kovac,\supit{g}
C.~L.~Kuo,\supit{l,m}
A.~E.~Lange,\supit{d}
E.~M.~Leitch,\supit{a}
M.~Lueker,\supit{d,h}
C.~B.~Netterfield,\supit{f}
H.~T.~Nguyen,\supit{d,h}
R.~W.~Ogburn~IV,\supit{l,m}
A.~Orlando,\supit{d}
C.~L.~Pryke,\supit{b}
C.~Reintsema,\supit{j}
S.~Richter,\supit{g}
J.~E.~Ruhl,\supit{n}
M.~C.~Runyan,\supit{d}
Z.~Staniszewski, \supit{d,h}
S.~Stokes, \supit{l,m}
R.~Sudiwala,\supit{c}
G.~Teply,\supit{d}
K.~L.~Thompson, \supit{l,m}
J.~E.~Tolan,\supit{l,m}
A.~D.~Turner,\supit{h}
P.~Wilson,\supit{h}
and C.~L.~Wong \supit{g}
\skiplinehalf
\supit{a}University of Chicago, KICP, 933 E.~56th St., Chicago, IL 60637 USA; \\
\supit{b}School of Physics \& Astronomy, University of Minnesota, 116 Church Street S.E., Minneapolis, MN 55455; \\
\supit{c}Dept.~of Physics and Astronomy, University of Wales, Cardiff,
CF24 3YB, Wales, UK; \\
\supit{d}California Institute of Technology, 1200 E.~California Blvd.,
Pasadena, CA 91125 USA; \\
\supit{e}Department of Physics \& Astronomy, University of British
Columbia, 6224 Agricultural Road, Vancouver, BC V6T1Z1, Canada; \\
\supit{f}Department of Physics, University of Toronto, Toronto, ON M5S
1A7, Canada; \\
\supit{g}Harvard-Smithsonian Center for Astrophysics, 60 Garden Street, Cambridge, MA 02138; \\
\supit{h}Jet Propulsion Laboratory, 4800 Oak Grove Dr., Pasadena, CA
91109, USA; \\
\supit{i}Service des Basses Tempratures, DRFMC, CEA-Grenoble, 17 rue des Martyrs, 38054 
Grenoble Cedex 9, France; \\
\supit{j}NIST Quantum Devices Group, 325 Broadway, Boulder, CO 80305, USA; \\
\supit{k}University of California, San Diego, La Jolla, CA 92093, USA; \\
\supit{l}Stanford University, Stanford, 382 Via Pueblo Mall, CA 94305,
USA; \\
\supit{m}Kavli Institute for Particle Astrophysics and Cosmology (KIPAC), Sand Hill Road 2575, 
Menlo Park, CA 94025, USA; \\
\supit{n}Physics Department, Case Western Reserve University, Cleveland, OH 44106 USA
}

\authorinfo{Send correspondence to C.~D.~Sheehy,\\
  E-mail: csheehy@uchicago.edu} 

\begin{document} 
  \maketitle 

\begin{abstract}
The Keck Array is a cosmic microwave background (CMB) polarimeter that
will begin observing from the South Pole in late 2010.  The initial
deployment will consist of three telescopes similar to BICEP2 housed in
ultra-compact, pulse tube cooled cryostats.  Two more receivers will
be added the following year.  In these proceedings we report on the
design and performance of the Keck cryostat.  We also
report some initial results on the
performance of antenna-coupled TES detectors 
operating in the presence of a pulse tube.  We find that the
performance of the detectors is not seriously impacted by the
replacement of BICEP2's liquid helium cryostat with a pulse tube cooled
cryostat.
\end{abstract}

\section{INTRODUCTION}
\label{sec:intro}  
The presence of a curl component in the polarization pattern of the
CMB is a generic prediction of inflation.\cite{Kamionkowski97,Seljak97}
A detection of 
this so-called B-mode of the CMB polarization would not only provide
``smoking gun'' style evidence that inflation occurred, but also offer
an important probe of GUT scale physics inaccessible to laboratory
experiments.  The Keck Array is optimized to detect the degree scale
peak of the inflationary B-mode spectrum, the amplitude of which
is parameterized by the ratio of tensor to scalar
perturbations in the early universe, $r \sim T/S$.  The simplest models
of single field, slow roll inflation favor values of $r\leq0.1$.  No B-mode
polarization has yet been detected.  The current best upper limit
on the B-mode spectrum of $r<0.24$ has been set by WMAP7 from
temperature anisotropy alone.\cite{Komatsu10}  To constrain $r$ further will
require sensitive measurements of the CMB polarization.  

BICEP2, which deployed to the South Pole in November 2009 and will
observe for the next two years, was designed with the explicit
goal of detecting inflationary B-modes.\cite{Ogburn10}  Its optical
design follows from BICEP1, which has constrained B-modes to
$r<0.72$.\cite{Chiang10}
BICEP2's small aperture (25 cm) and on-axis, cold, refracting optics
are ideal for characterizing and minimizing instrument systematics.
The optics, which consist of
anti-reflection coated lenses and infrared blocking filters, are 
optimized for a single frequency.\cite{Aikin10}  The focal plane
consists of 512 antenna-coupled, TES bolometers operating at 150 GHz,
and are 
described in detail in Ref.~\citenum{Kuo08}.  Each detector is sensitive to a
single polarization orientation.  Differencing the signal between
members of
256 pairs of spatially coincident detectors allows for a high level of
rejection of common mode noise.  
The initial performance of the
BICEP2 detectors is described in Ref.~\citenum{Brevik10}.

\begin{figure}
  \begin{center}
    \begin{tabular}{c}
      \includegraphics[height=6cm]{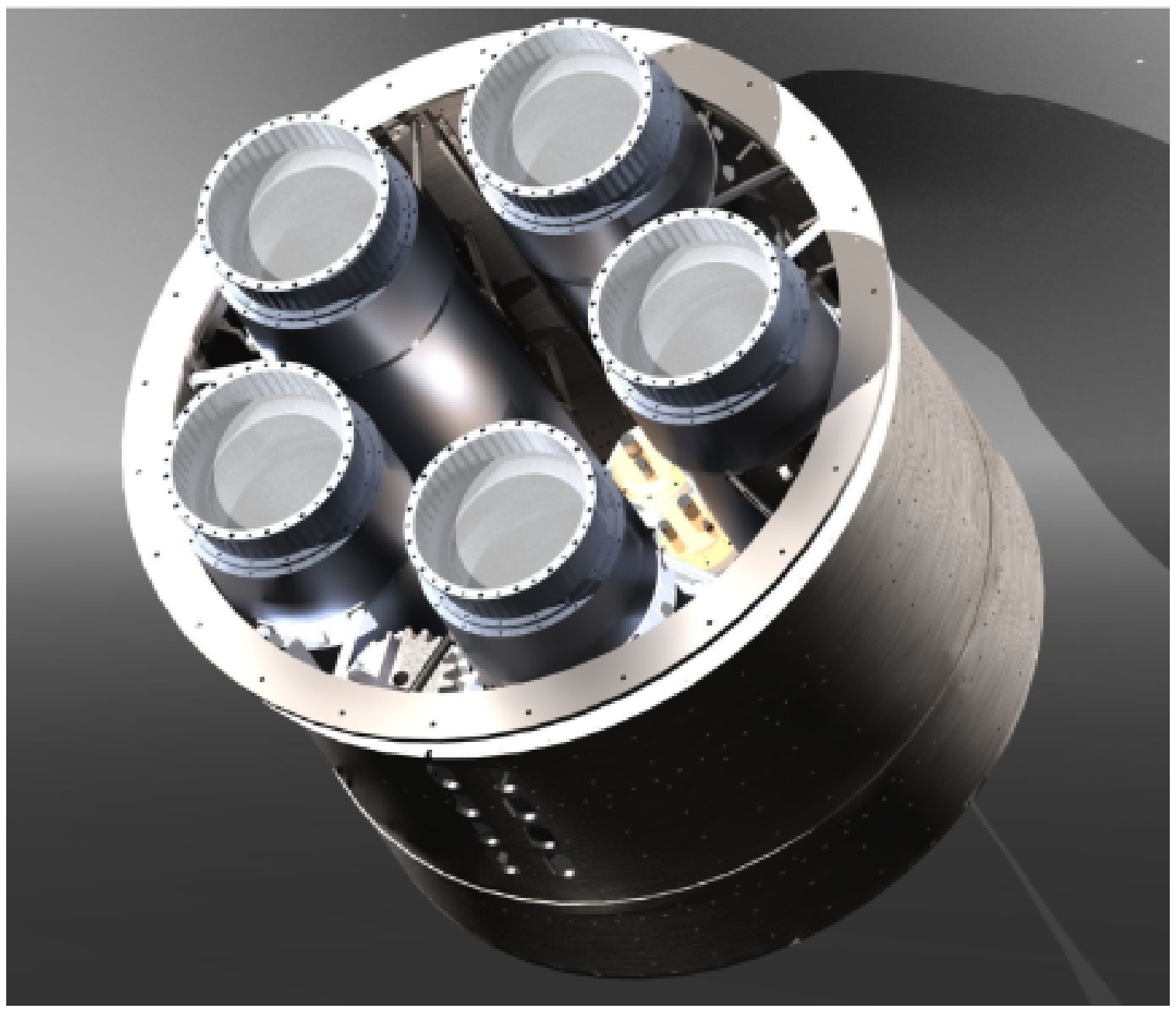}
      \includegraphics[height=6cm]{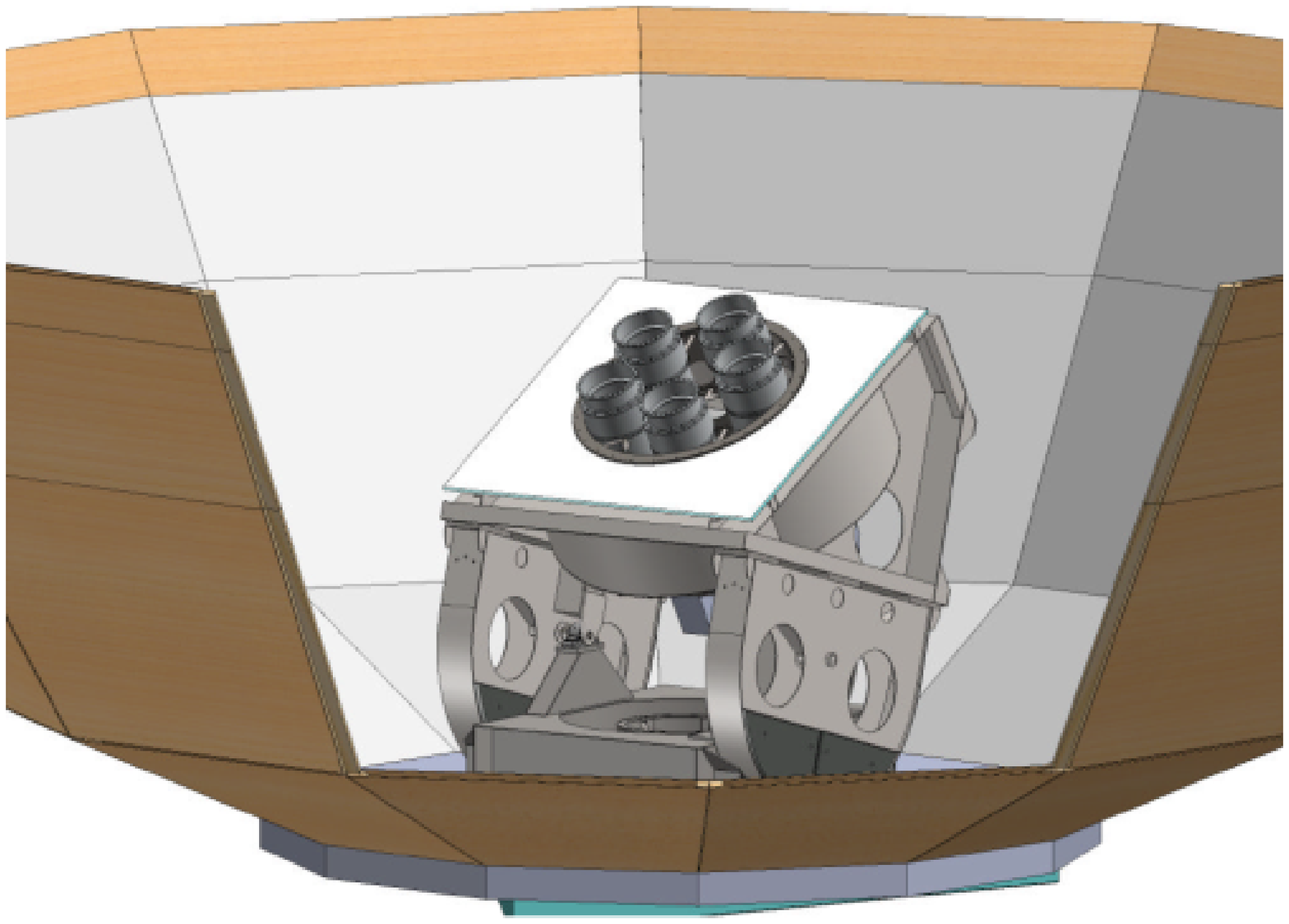}
    \end{tabular}
  \end{center}
  \caption[example] 
	  { \label{fig:drum} 
Left: CAD rendering of the modified drum of the DASI mount with 5
  Keck cryostats installed.  Right: Keck cryostats installed in the
  DASI mount and ground shield (cutaway).
}
\end{figure}

Detecting the B-mode polarization of the CMB will require
unprecedented sensitivity.  To achieve this, the Keck Array will
deploy five BICEP2 style receivers, beginning with three in November
2010 and following with two more in late 2011.  These receivers
will be fit into an existing mount at the South Pole originally built
for the DASI experiment and most recently used for the QUAD
experiment.\cite{Kovac02,Pryke09}   The azimuth/elevation mount has a drum
which can rotate about its axis, thus duplicating BICEP1 and BICEP2's
ability to rotate the entire optics around the boresight.  The mount
with five receivers installed is shown in Figure \ref{fig:drum}.  Because of
the space constraints imposed by the existing infrastructure,
maximizing the number of receivers requires designing a very compact
cryostat.  For this reason, and because the South Pole station does
not have the resources to provide enough liquid helium to keep many
receivers cold through the 9 months during which the station is
inaccessible, the Keck Array will use pulse tube cooled cryostats in
place of the liquid cryogen cryostats of BICEP1 and BICEP2.

Section \ref{sec:cryostat} discusses the design of the Keck cryostat and
its thermal performance.  Section \ref{sec:performance} discusses the
initial performance of BICEP2 detectors operating in the cryostat,
paying specific attention to any possible degradation in performance
compared to BICEP2.

\section{CRYOSTAT DESIGN}
\label{sec:cryostat}

\begin{figure}
  \begin{center}
    \begin{tabular}{c}
      \includegraphics[height=4.7in]{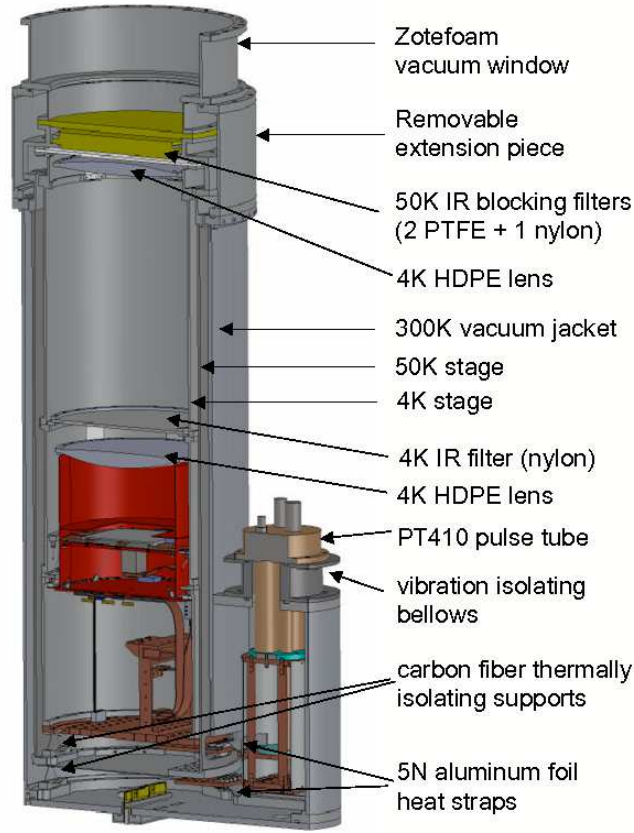}
    \end{tabular}
  \end{center}
  \caption[example] 
	  { \label{fig:cryostat_xsec} 
Cross section of the Keck cryostat.
}
\end{figure}

The Keck cryostat, shown in cross section in Figure
\ref{fig:cryostat_xsec}, uses a Cryomech PT410 two-stage pulse tube
cooler, mounted on vibration isolating bellows.  According to
Cryomech's published cooling capacity curves, the first stage of
the PT410 will maintain a temperature $T<50$ K while conducting as
much as 45 W of power.  This substantial amount of cooling power allows us to
thermally sink three separate infrared blocking filters to the 50 K
stage of the pulse tube.  In BICEP2 these filters are mounted to
liquid helium vapor cooled stages.  
Also according to the system specifications given by Cryomech, the 2nd
stage of the PT410 
maintains a temperature of $\sim3.5$ K at $0.5$ W of conducted power,
rising to $\sim4.25$ K at $1.0$ W of power, nearly independent of the
power on the 1st stage.  We have verified this in the lab for a
nominal loading of $25$ W on the 1st pulse tube stage.  Because of the
finite cooling capacity of the PT410 we have taken great care in the
design of our mechanical supports to
minimize parasitic loads onto the 4K stage, and in the design of our
heat straps to maximize their thermal conductivity.

A removable cryostat extension encloses the filter stack and the top
of the optics tube.  This extension could be swapped out for one of a
different height, providing flexibility in filter design and allowing
the possible addition of a stepped half wave plate.

The lack of a tank for liquid cryogens has allowed us to highly
compactify the cryostat.  The largest diameter of the BICEP2/Keck insert is
14.76'' and the main body of the Keck cryostat has an outer diameter of
$18.0$'', leaving very little clearance between stages.  
The largest diameter of the cryostat, not including the off-axis
``doghouse'' housing the pulse tube, is the 19.5'' diameter of the
extension piece and its mating flange.

Magnetic shielding inside the cryostat is a sheet of rolled Amuneal A4K,
which we have clamped
to the 50 K cylinder wall, drilled in place (mostly 
through pre-drilled holes in the shielding), and attached with low
profile screws.  The two ends of the A4K shield overlap by a few
inches.  The geometry of this shield is different from BICEP2, which
is a rolled and welded cylinder.
Also, because Amuneal has phased out Cryoperm and replaced it with A4K,
which is supposedly both higher performance and cheaper,
the use of A4K in the Keck cryostat is a difference from BICEP2.

\subsection{THERMALLY ISOLATING MECHANICAL SUPPORTS}
\begin{figure}
  \begin{center}
    \begin{tabular}{c}
      \includegraphics[height=2.5in]{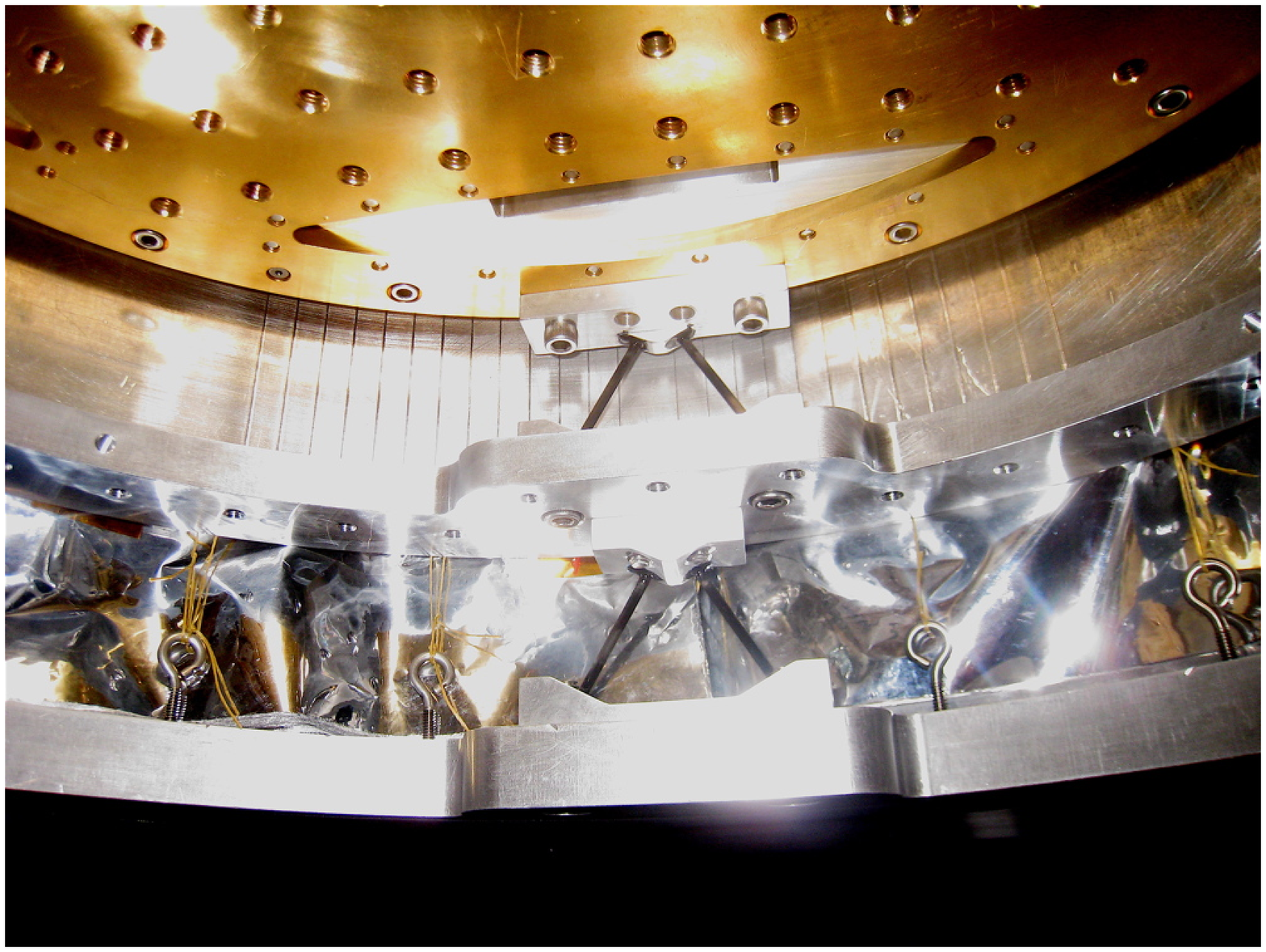}
      \includegraphics[height=2.5in]{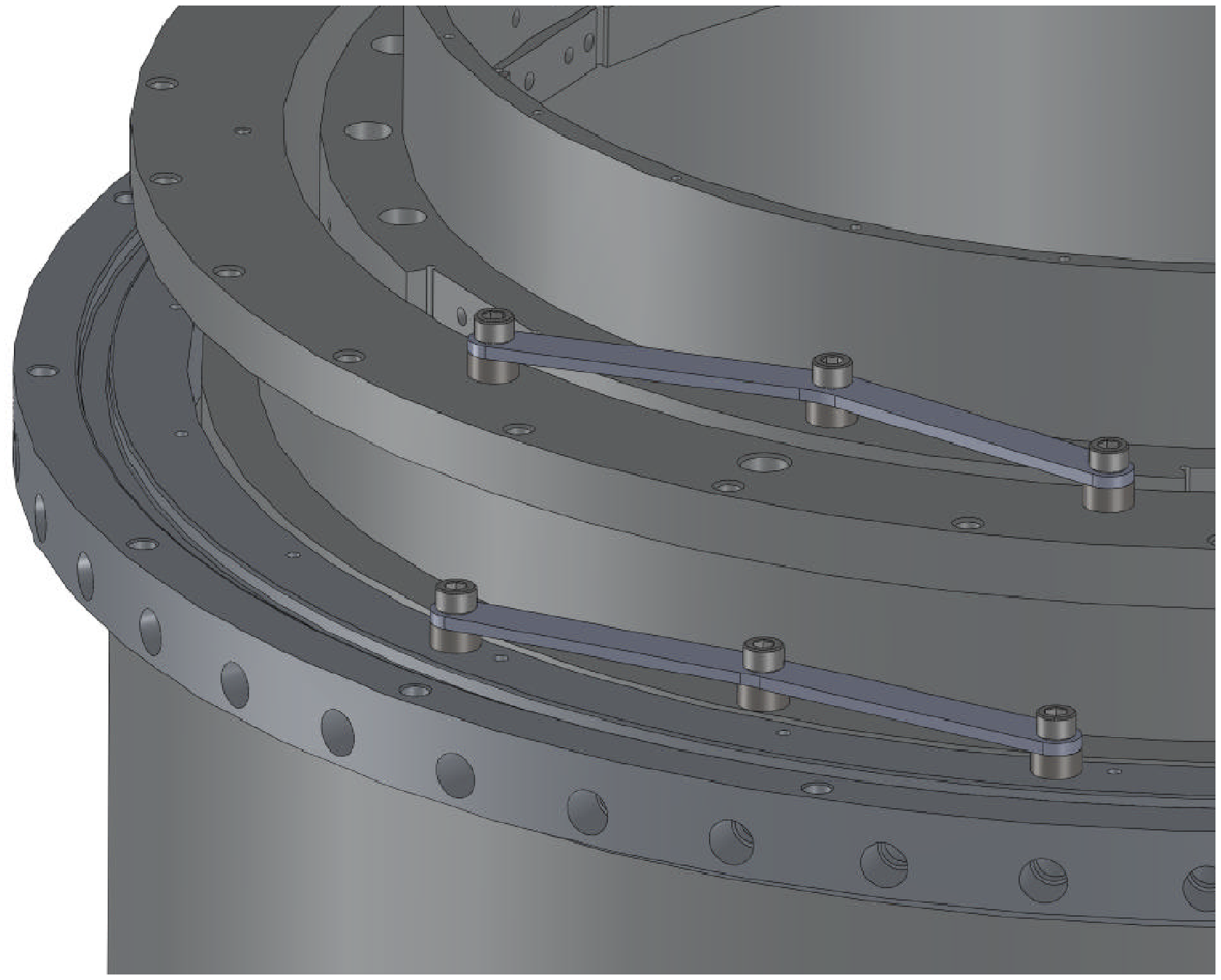}
    \end{tabular}
  \end{center}
  \caption[example] 
	  { \label{fig:supports} 
Left: Carbon fiber mechanical supports at the back end of the cryostat.  The
  bottom standoff is one of three two-member trusses thermally
  isolating the 50 K assembly from the 300 K vacuum jacket.  The top
  standoff is one of the three trusses supporting the 4 K assembly
  off of the 50 K assembly.  Right: Titanium 6-Al-4V mechanical
  supports at the top of the cryostat.  These constrain the top of the
  cold stages in the radial direction while allowing a large amount of
  differential thermal contraction in the axial direction.
}
\end{figure}

Each cold stage is mechanically supported at the back end of the
cryostat by three sets of two-member carbon fiber trusses.  These
trusses are shown in the left hand panel of Figure \ref{fig:supports}.
Each truss 
consists of two $.108$'' diameter by $2.084$'' length carbon fiber rods
epoxied into $.141$'' diameter by $.310$'' deep holes drilled into
aluminum blocks.

The properties of carbon fiber at temperatures of 4 K and below were
investigated in detail by Ref.~\citenum{Runyan08}.  The ratio of elastic
modulus to thermal conductivity of carbon fiber is very high at
cryogenic temperatures.  At 4 K this ratio is higher than any other
other material tested in Ref.~\citenum{Runyan08}, including G10 and Vespel.
At 50 K this ratio 
drops but is still comparable to G10.  The successful use of carbon
fiber as thermally isolating truss supports for the sub-Kelvin
stages of the SPIDER and BICEP2 focal planes motivated its use in the
Keck cryostats.  FEA analysis of the 
carbon fiber trusses indicates they will provide sub-arc-minute
pointing accuracy and a mechanical strength adequate for a 300 lb load
with a safety factor greater than five.  We have not measured their
thermal conductivity, but a simple calculation predicts of order $16$ mW
conducted from the 50 K to 4 K assembly. 

The cold stages of the cryostat are supported at the front end by
three v-shaped titanium 6-Al-4V ``boomerangs,''
shown in the right hand panel of Figure \ref{fig:supports}.
These pieces are very stiff in the radial direction but quite flexible
along the cryostat's axis, an acceptable configuration because of the
stiffness of the rear carbon fiber supports in the axial direction.
The flexibility of the titanium supports in the axial direction safely
allows a large ($~0.2$'') differential thermal contraction
between the 
300 K and 50 K stages.  Titanium was chosen for its high stiffness
and for its large yield-strength to thermal conductivity ratio at
low temperatures, which was needed to accommodate the radial
differential thermal contraction between stages.  The total loading
onto 4K from the titanium supports is estimated from a simple
calculation to be $35$ mW.

\subsection{ALUMINUM FOIL HEAT STRAPS}
\begin{figure}
  \begin{center}
    \begin{tabular}{c}
      \includegraphics[width=5in]{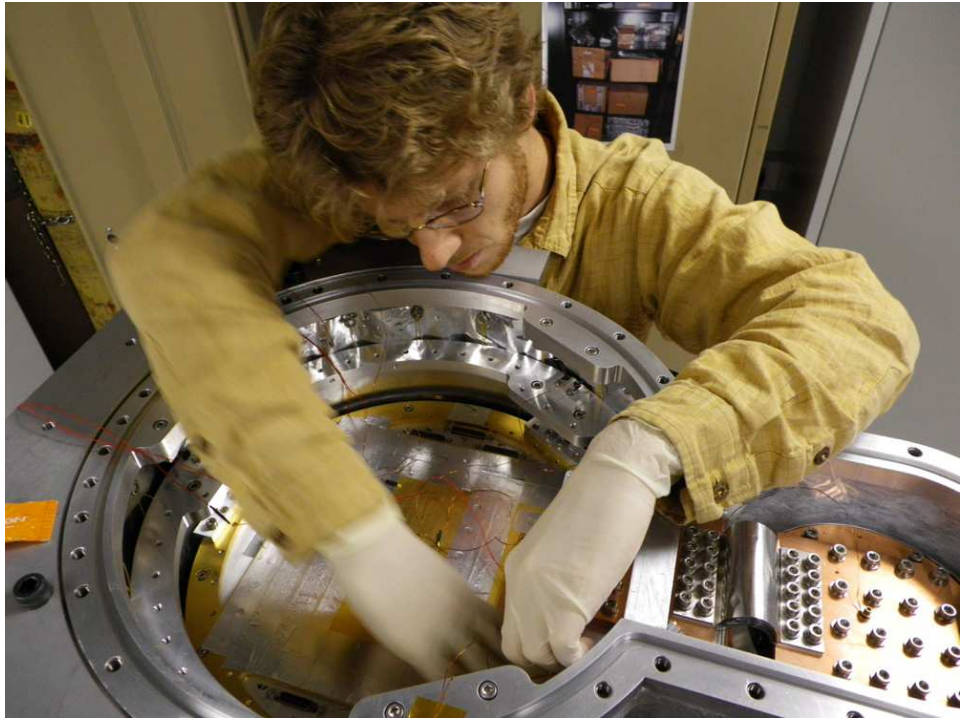}
    \end{tabular}
  \end{center}
  \caption[example] 
	  { \label{fig:heatstrap} 
The back of a Keck cryostat with the vacuum close-out plate removed,
  showing the 300 K, 50 K, and 4 K stages.  The 50 K aluminum foil heatstrap
  is visible in the lower right corner.
}
\end{figure}

The liquid helium tank in the BICEP2 cryostat provides a large surface
on which to directly mount the 4K receiver.  The Keck cryostat has no
such surface.  As such we
have designed a very high thermal conductivity heat strap to
connect the pulse tube cold head to the BICEP2 receiver.  Running
contrary to this goal is the need for the heat strap to be flexible in
order to avoid transmitting vibrations from the pulse tube to the
focal plane.  To satisfy both these competing criteria we have
constructed our heat straps from stacks of 99.999\% pure (5N) aluminum
foil.  The stacks of foil are pressed against OFHC copper blocks at
either end via an aluminum clamping plate and 10 18-8 stainless steel,
1/4''-20 bolts tightened to a torque of $\sim10$~ft-lbs.
(We exceed the torque
limit for the bolts because we are using nuts and not threaded holes.)
The 50 K heat
strap is shown in Figure \ref{fig:heatstrap}.  The strap is very
flexible along the axis connecting the pulse tube cold head and the 4K
mating surface, as well as along the axial direction of the cryostat,
and will transmit motion in these directions very weakly.

Great care is needed in the assembly of the heat straps to ensure good
thermal contact between foils.  To this end we have constructed an
assembly jig  
that holds the foils against the copper blocks via the aluminum
pressure plate during drilling.  The pre-clamping of the foils prevents
any edges from being raised.  The pressure plate is then fastened in
place with screws and nuts, and the heat strap can be removed from the
jig.  

The use of 5N Al was motivated by its very high thermal conductivity
at 4 K, which is approximately a factor 10 higher than OFHC
copper with a residual resistivity ratio of
RRR$=50$,\cite{Woodcraft05,Ekin06} and by its low price and availability 
relative to 
5N Cu.  A very low cross sectional area is needed to
achieve an acceptable temperature rise, thus allowing us to maximize the 
flexibility of the strap.  However, the use of high purity aluminum in
cryogenic applications seems to be discouraged in the
literature.\cite{Woodcraft05}  The reason cited is that the oxide layer on aluminum
gives it a very high thermal contact resistance, rendering it
unsuitable for use unless extraordinary measures are
taken.  Although we have taken care to maximize
the thermal contact between foils, the foils are not welded.
Because the foils are stacked 
they mate to their end blocks in parallel.  Thus the conductance of the straps
is limited by the contact resistance between foils.
Nonetheless, we have not observed unacceptably high contact
resistances.  A 4 K strap consisting of 10 foils, each $3.00$'' wide by
$\sim2.5$'' free length by $.004$'' thick, was measured to have a
conductance of $1.1$ W/K at 4 K, with a temperature rise of $0.48$ K
across it while conducting $0.53$ W of power.
Given the range of thermal conductivities among different samples of
5N Al, which is given by Ref.~\citenum{Woodcraft05} to be
$4-20\times10^3$~W/m/K,
our calculations imply that the contact conductance between foils at
$4$~K must be somewhere in the range $3-9$~W/K.  
We have since added an
identical 10 foil stack to the reverse sides of the strap to
increase its conductance by a factor of two.  
The thermal
contact resistance of high purity aluminum does not appear to be an
insurmountable barrier to its use in high conductivity, flexible heat
straps.  
To increase 
conductance further it would be possible to construct the end
blocks from high purity aluminum (if not 5N Al then perhaps Al 1100)
and e-beam weld the foils to the blocks.

\section{PERFORMANCE}
\label{sec:performance}
The major difference between BICEP2 and Keck is the use of a pulse
tube cooled cryostat in place of a
liquid helium cooled cryostat.  We
have performed initial commissioning of BICEP2 style detectors in the
Keck cryostat.  We have not yet observed any large differences in
performance of the detectors while running them in the presence of a
pulse tube as compared to BICEP2.

\subsection{DETECTOR NOISE}
\begin{figure}
  \begin{center}
    \begin{tabular}{c}
      \includegraphics[width=6.5in]{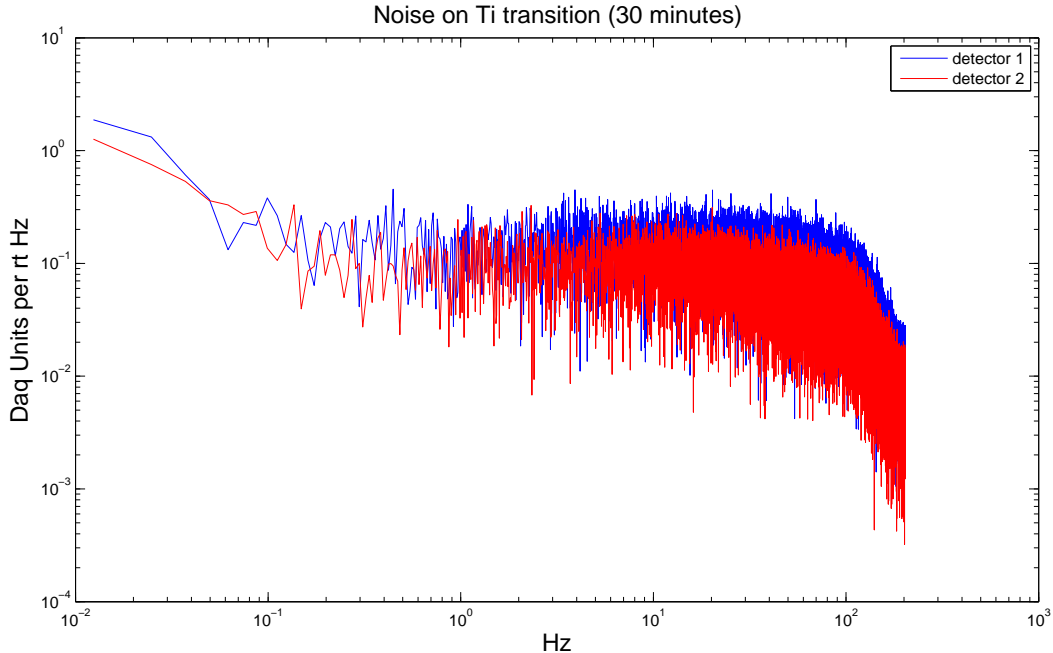}
    \end{tabular}
  \end{center}
  \caption[example] 
	  { \label{fig:noisepsd} 
Noise PSDs of two detector pixels, which were
  shaded to allow them to be biased onto transition in the radiation
  environment of the lab.  In neither curve is the $1.4$~Hz pulse tube
  frequency visible.
}
\end{figure}

One of the primary concerns regarding the performance of Keck as compared
to BICEP2 is the possibility of elevated detector noise at the
$1.4$~Hz pulse tube frequency.
We have measured the noise power spectral density (PSD) of two
detector pixels, which are shown in Figure~\ref{fig:noisepsd}. 
A zotefoam vacuum window was in place on the cryostat and both 
pixels were shaded to allow the titanium TESs to be biased onto
transition in the elevated loading conditions of the
lab.\cite{Orlando10}   
The sampling rate was $406$~Hz.  The PSDs shown are an average of 10
separate PSDs of $2^{16}$ samples each, for a total integration time of
$27$~minutes.  
In neither of the PSDs is the $1.4$~Hz pulse tube frequency visible.


\subsection{MAGNETIC PICKUP}
There are differences between BICEP2 and Keck that might affect
the response of the SQUID multiplexers to an external magnetic field.
First, as discussed in Section~\ref{sec:cryostat}, the magnetic
shielding in the first Keck cryostat is of a slightly 
different geometry and size, and is made of a different material than
that in BICEP2.  Second, the
pulse tube in the Keck cryostat introduces a large piece of stainless
steel near the detectors. 

\begin{figure}
  \begin{center}
    \begin{tabular}{c}
      \includegraphics[height=2.5in]{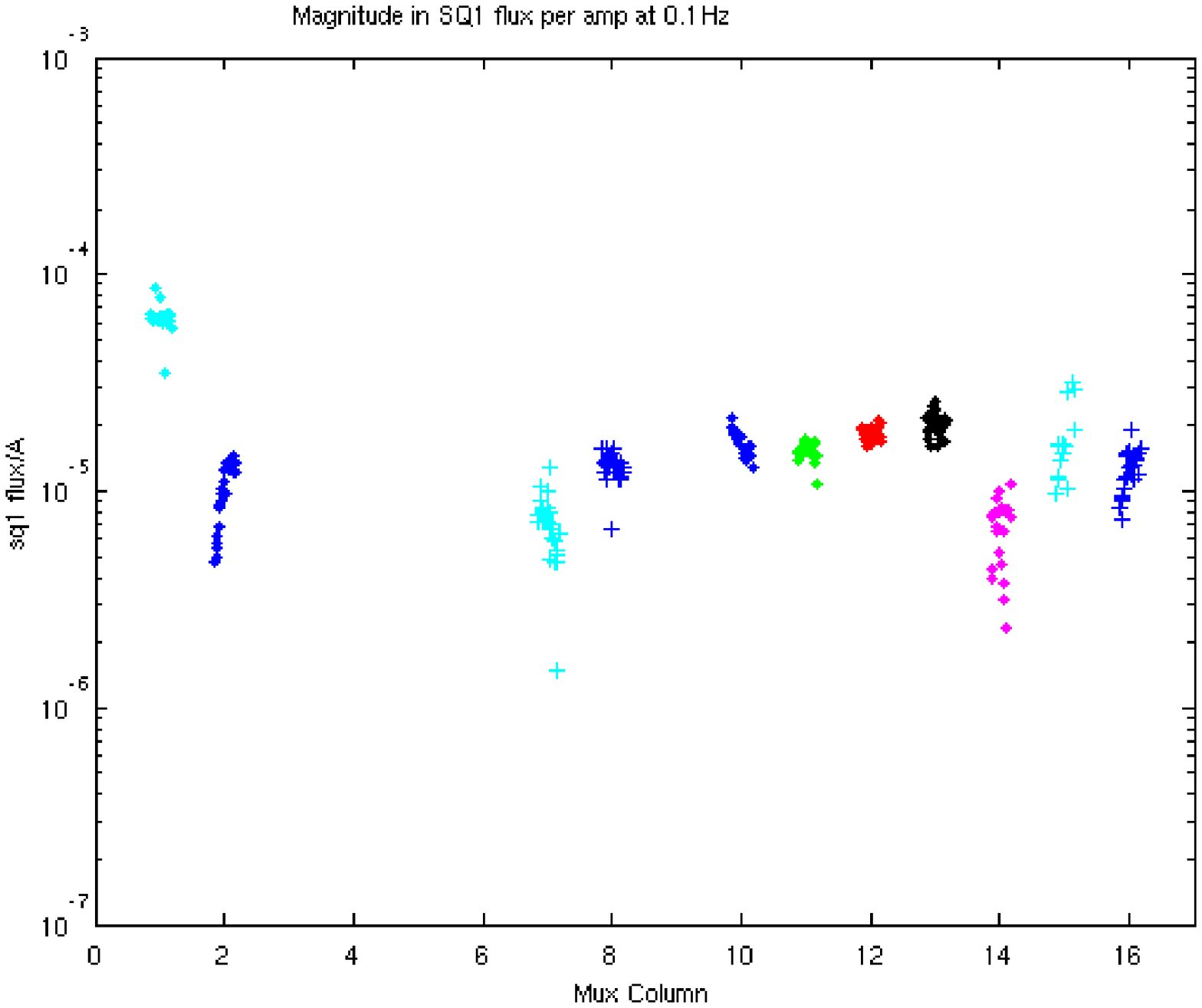}
      \includegraphics[height=2.5in]{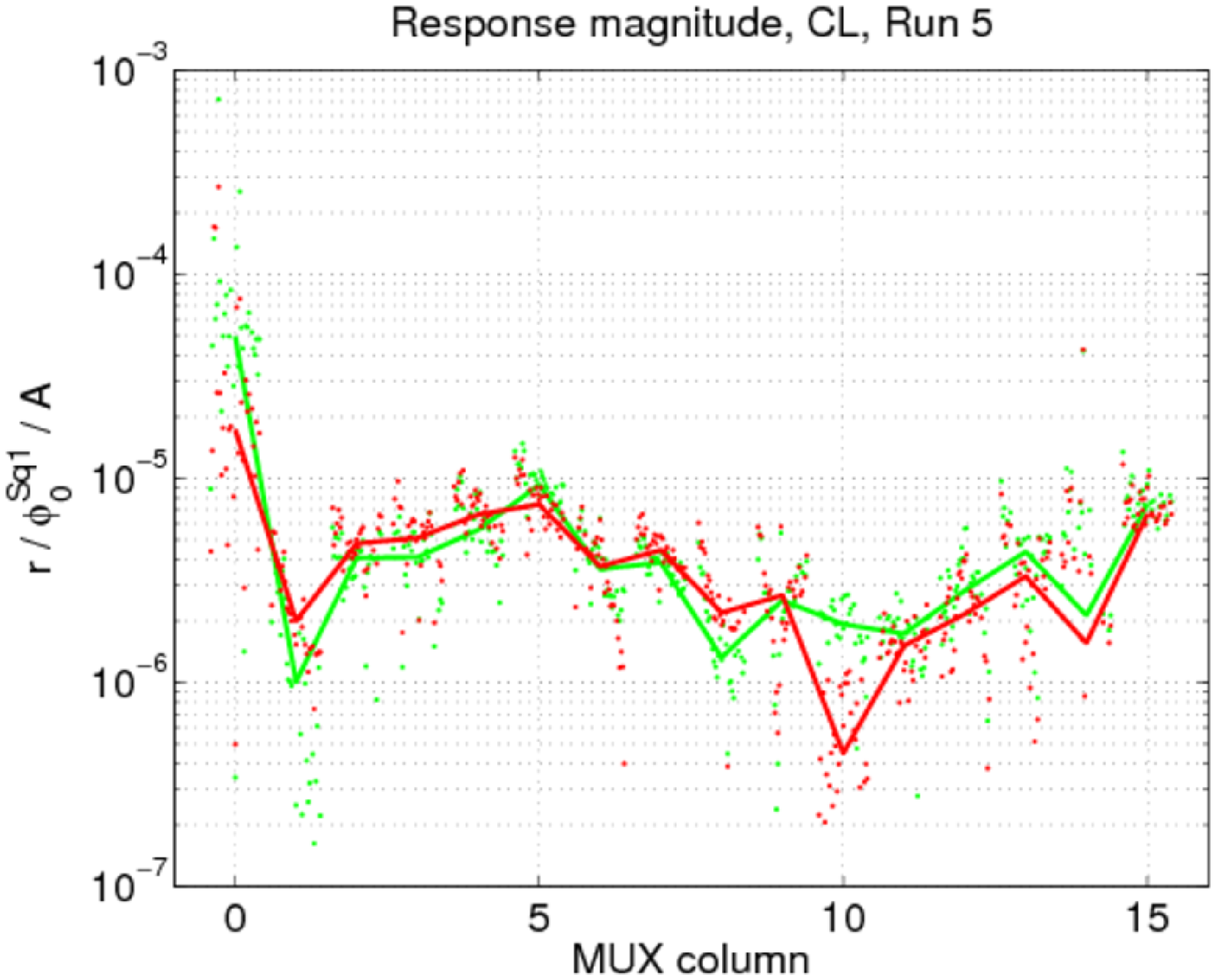}
    \end{tabular}
  \end{center}
  \caption[example] 
	  { \label{fig:magpickup} 
Left: SQUID response to a 0.1 Hz magnetic field in units of SQ1
  $\phi_0$ per Amp peak-to-peak for 
  the Keck cryostat, normalized to be comparable to a similar test
  performed for BICEP2.  Right: The analogous test done for BICEP2.
}
\end{figure}

We have tested the magnetic response of the SQUID multiplexers to
a magnetic field induced by Helmholtz coils placed around the
cryostat.  The coils have $r=0.5$ m and $N_{turns}=18$, operated
with a current of 5.8 Amps peak-to-peak, and sweeping in frequency from
0.05 to 100 Hz.  Figure \ref{fig:magpickup} shows the response of the
SQ1 arrays operating closed-loop to the 0.1 Hz magnetic field, as well
as the response from a 
similar test conducted for BICEP2.  (Note that at the time this test
was performed the Keck magnetic shielding was Cryoperm mounted to the
4 K assembly.  This has subsequently been changed to A4K mounted at 50
K.  Also, the focal plane used in the BICEP2
run uses single turn series SQUID arrays, while the Keck run has three
turn SSAs.)  There is a factor $\sim3$ increase in magnetic response of
the Keck SQUIDs as compared to BICEP2.  Another effect not shown
here is that the Keck SQUID response does not appear to fall off with
frequency as compared to 
BICEP2.  However, we
do not believe that either of these modest 
changes in SQUID response will significantly impact detector
performance.  
Any residual magnetic response must be removed in analysis,
which is discussed in detail in Ref.~\citenum{Ogburn10}.

\section{CONCLUSION}
The Keck Array will deploy 3 BICEP2 style receivers to the South Pole
in November 2010.  The initial deployment will likely be three
receivers at 150 GHz.  Two more receivers will be added in November
2011.  We have designed and tested ultra-compact,
pulse tube cooled cryostats.  The cryostat design allows the maximum 
possible number of receivers to be fit into the existing DASI mount at
the South Pole station.  Initial detector commissioning does not reveal
any major impact on their performance arising from the new cryostats
or the pulse tube,
both of which constitute the main difference between Keck and BICEP2.
Because 
of this, we expect the Keck Array to offer unprecedented sensitivity to
the B-mode component of the CMB polarization.

\acknowledgments 
The Keck-Array is funded by the National Science Foundation through
grants ANT-0742592 and ANT-0742818 and by the Keck Foundation. We
acknowledge assistance from the KICP at the University of Chicago through
the grant NSF PHY-0114422, and the support of the NASA Postdoctoral
Program for Zak Staniszewski.  This paper is dedicated to the memory of
Andrew Lange, who died under tragic circumstances in January 2010.
Andrew made invaluable 
contributions to the field of experimental cosmology.  He played a central
role in the conception of the BICEP2 and Keck experiments.  His scientific
aptitude, leadership, and unique abilities to recognize and develop young
scientists are sorely missed by his friends and colleagues.

\bibliography{sheehy}   
\bibliographystyle{spiebib}   

\end{document}